\documentclass[a4paper,12pt]{article}

\usepackage[dvips]{epsfig}

\usepackage{amsfonts}
\usepackage{amssymb}
\usepackage{amscd}
\usepackage{amsthm}
\usepackage{latexsym}
\usepackage{amsbsy}

\def\a{\alpha}
\def\b{\beta}
\newcommand{\g}{\gamma}
\def\d{\delta}

\newcommand{\ep}{\epsilon}

\newcommand{\la}{\lambda}

\newcommand{\si}{\sigma}

\def\cH{\mbox{${\cal H}$}}

\def\dim{{\rm dim}}

\def\det{{\rm det}}

\def\tr{{\rm tr}}

\def\exp{{\rm exp}}
\def\tr{{\rm tr}}
\def\Tr{{\rm Tr}}

\newcommand{\Bt}{\tilde B}
\newcommand{\Mt}{\tilde M}

\def\uqt{$U_q(su(2))$}

\def\llap#1{\hbox to 0pt{\hss#1}}
\def\pola{a\llap{\hbox{\char'30\kern-1.2pt}}}
\def\pole{e\llap{\hbox{\char'30\kern-.8pt}}}

\newcommand{\non}{\nonumber\\}

\newcommand{\lan}{\langle}
\newcommand{\ran}{\rangle}

\newcommand{\beq}{\begin{equation}}
\newcommand{\eeq}{\end{equation}}
\newcommand{\beqa}{\begin{eqnarray}}
\newcommand{\eeqa}{\end{eqnarray}}
\newcommand{\barr}{\begin{array}}
\newcommand{\earr}{\end{array}}
\newcommand{\ben}{\begin{enumerate}}
\newcommand{\een}{\end{enumerate}}
\newcommand{\bit}{\begin{itemize}}
\newcommand{\eit}{\end{itemize}}

\newcommand{\refeq}[1]{(\ref{#1})}

\def\R{{\mathbb R}}

\def\trr{\triangleright}

\def\rep{representation }
\def\reps{representations }

\def\cF{{\cal F}}
\def\cG{{\cal G}}

\def\cR{{\cal R}}

\def\cG{{\cal G}}

\def\nn{\nonumber}

\def\tens{\otimes}

\def\mg {{\mathfrak g}}

\def\eps{\varepsilon}

\def\cH {{\cal H}}

\def\wh{\hat w}

\begin{document}
\begin{flushright}
LMU-TPW 03-04 \\
IFT-03/12\\

\end{flushright}
\begin{center}
{\Large \bf Algebraic brane dynamics on SU(2):\\ 
excitation spectra\footnote{Work supported   
by Polish State Committee for Scientific Research (KBN) under contract
2 P03B 001 25  (2003-2005)}}\\[20pt]
J. Pawe\l czyk$^a$\footnote{Jacek.Pawelczyk@fuw.edu.pl}
and  
H.\  Steinacker$^b$\footnote{Harold.Steinacker@physik.uni-muenchen.de}\\[2ex] 
{\small\it ${}^a$Institute of Theoretical Physics\\ Warsaw University,
   Ho\.{z}a 69, PL-00-681 Warsaw, Poland\\[1ex]

    ${}^b$Sektion Physik der Ludwig--Maximilians--Universit\"at M\"unchen\\
       Theresienstr.\ 37, D-80333 M\"unchen  \\[1ex] }

\vspace{1.5cm}

{\bf Abstract} \\

\end{center}

We analyze the dynamics of $D2$--branes on $SU(2)$
within a recently proposed matrix model, 
which works for finite radius of $SU(2)$. 
The spectrum of single-brane excitations turns out to
be free of tachyonic modes. It is similar to the spectrum found
using DBI and CFT calculations, however the triplet of rotational 
zero modes is missing. This is attributed to a naive treatment 
of the quantum symmetries of the model. 
The mass of the lightest states connecting two
different branes is also calculated, and found to be proportional
to the arc length for small angles.

\noindent
 
\vspace{1.5cm}

\newpage

\section{Introduction}

In recent years  the structure of $D$-branes in a WZW models
has attracted much attention. The background of the compact Lie groups
$G$ is known to
 carry a nontrivial (NSNS) $B$ field which is not closed.  
It has been shown, using CFT \cite{as,gawedzki} and  DBI (Dirac-Born-Infeld)
\cite{jbds} descriptions, that stable branes can 
wrap certain conjugacy classes in the group manifold. The problem of
the brane 
description has also been attacked from a different perspective, by
the matrix model \cite{myers} of D0-branes. The latter approach was supported 
 by CFT calculations 
\cite{ars2} and led to a beautiful picture where, in a 
special limit, the macroscopic branes could be viewed as a
bound state of $D0$-branes. Finally, the spectra of branes in WZW models were
calculated using K-theory \cite{tw-K,mms,moore-rev}.

These various approaches focus on different aspects of D-brane
physics, and have
different range of applicability.
For example, the DBI approach is valid only in the large $k$
limit, while the standard matrix model 
can handle only a certain subset of ``small''
branes. 
CFT is the most complete description, 
but it lacks the physical picture of the brane,
and involves too many degrees of freedom.

Attempting to reconcile these various approaches, we proposed in
\cite{PS1} a matrix description of $D$ - branes on 
$SU(2)$ based on quantum symmetries.
This led to a quantum algebra of (noncommutative) matrices, which
reproduces all static properties of all stable
$D$-branes on $SU(2)$, 
including finite size effects. 
This was extended in \cite{PS2} to cover all
untwisted branes for higher groups such as $SU(N)$.
In these papers, we were concerned only with static properties 
of the branes, ignoring the brane dynamics.
This is the problem we want to address here, attempting to describe the 
dynamics of the branes in terms of gauge theories based on our
quantum algebra.

In the present paper we calculate the spectra of excitations of
branes within this matrix model. 
Because the paper is quite technical and the results are easy to state 
we present them here. The resulting 
excitation spectrum for a single brane 
$D_{2\la},\;\la\in\{0,..k/2\}$ is (in the
large $k$ approximation)  
\beqa\label{mass-s}
m^2 &=& \a^2\; j^2, \qquad\quad  j = 1, 2, ..., 2\la \nn\\
m^2 &=& \a^2\; (j+1)^2,  \quad j = 0, 1, ..., 2\la,
\eeqa
where $\a\sim r/k$ and $r$ is the radius of $SU(2)$
(which should be $r \sim \sqrt{k}$).
The spectrum \refeq{mass-s} is very close to the known results \cite{jbds}.
In particular there is no tachyon, which 
shows that the model is quite reasonable as the branes are stable. 
Unfortunately, \refeq{mass-s} does not contain
the massless modes corresponding to the freedom 
of rotating branes inside $S^3$, which it should. 
We shall discuss this point at length in the
paper.

\begin{figure}[htpb]
\begin{center}
   \epsfbox{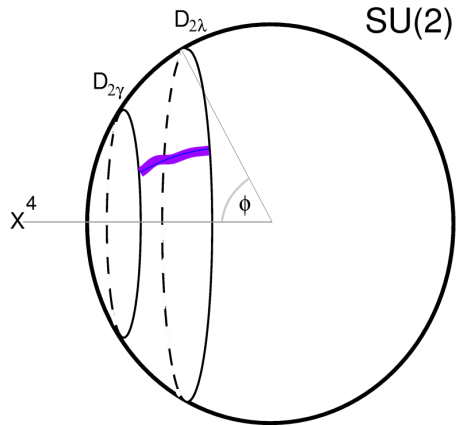}
\end{center}
 \caption{D2-branes on $SU(2)$, with string. The angle is 
$\phi = \frac{(2\la+1)\pi}{k+2}$.}
\label{fig:2branes}
\end{figure}

Furthermore,  the lightest states connecting two 
different (parallel) 
branes $D_{2\la}$ and $D_{2\g}$ have masses
\beq
m^2=4 r^2 \sin^2\frac{(\la-\g)\pi}{k+2}.
\label{mass-long}
\eeq
This formula  
gives $4 r^2 (\frac{(\la-\g)\pi}{k+2})^2$ 
in the large $k$ approximation,  
which is  the arc length of the string stretched
between the two branes. This is in perfect agreement with stringy intuition and
results obtained by other means 
\cite{moore-rev}.
For finite $k$ 
and large angles, the result deviates from this simple
geometrical interpretation, and it would be quite interesting 
to verify this.
The scenario is sketched in Figure \ref{fig:2branes}.

We should point out that the lack of rotational zero-modes
in this model is presumably due to an oversimplification in the way we
treat the model. In order to keep things as simple as possible, 
we have considered our action as an ordinary, classical
matrix model. We will discuss later that this is inconsistent
with the (thus far formal) quantum group symmetries, so that they do not
guarantee zero modes as they should. Hence this work should be 
seen as a first step towards a fully consistent treatment.
Nevertheless, we believe that the models 
are very interesting as they are, being perhaps 
the first matrix models for branes on a curved background.

The detailed calculations leading to the above results are 
presented in Section \ref{single} where we 
analyze the spectra of single branes, and in Section \ref{inter} where we 
consider excitations of two parallel
branes. These two sections form the heart of the paper. They are preceded by
an introduction of the basic variables and equations 
we are going to use. We close the paper with a discussion of the results, and 
several appendices containing technicalities.

\section{Branes in the SU(2) WZW model}

In this section we recall the necessary facts about D-branes in
the $SU(2)$ WZW
model at level $k$. It is based on results obtained in several papers 
\cite{as,FFFS,jbds,mms}, and contains
no new material.
 
We know that there is only a discrete 
set of stable $D$--branes on $G=SU(2)$ (up to global rotations),
one for each integral weight\footnote{Here and in the following $\la$ 
will denote the (half-integer) spin, hence the weight is $2\la$.}
 $2\la\in P_k^+=\{0,1,\dots k\}$.
They are given by the conjugacy class 
$D_{2\la} = \{g^{-1}t_{2\la}\ g\} $ of the SU(2) element
\beq
t_{2\la} = \exp(2 \pi i \frac{H_{2\la} + H_\rho}{k +2})=\left(
\begin{array}{cc}q^{2\la+1}&0\\0&q^{-(2\la+1)}
\end{array}\right)
\label{stable-branes}
\eeq
where $\rho = \frac 12  \a$  is the Weyl vector ($\a$ is the only
root),
\beq
q = e^{\frac{i \pi}{k+2}}
\eeq
 and 
$H_\la$ is the Cartan generator 
corresponding to $\la$.
If one represents  $SU(2)$ as a
3-sphere of radius $r$  
embedded in $\R^4$, then the (classical) position of the brane is given 
by 
\beq\label{position}
x^4= \frac r2\; \tr (t_{2\la})=r\;\cos\frac{(2\la+1)\pi}{k+2}.
\eeq
It is known that $r^2=k$ for large $k$.
Hence the  world-volume of the
(D2-) brane looks like a 
2-sphere embedded in $S^3$ at the angle
$\frac{(2\la+1)\pi}{k+2}$, 
see  Figure \ref{fig:2branes}. 

The fluctuations of these branes have been analyzed in the first paper of 
\cite{jbds}.
It turns out that they decompose as 
$1\otimes \la\otimes\la=(\oplus_{j=0}^{2\la-1}\ j)
\oplus(\oplus_{j=1}^{2\la}\ j)\oplus(\oplus_{j=1}^{2\la+1}\ j)$. The middle
series is gauged away. 
The spectrum of the physical fluctuations is 
\beqa
m^2_j= \frac 1{k }\left\{ 
\begin{array}{ll} (j+1)(j+2)  &\qquad \mbox{for } j=0,1,...2\la-1\\
(j-1)j&\qquad \mbox{for } j=1,2,...2\la+1  \end{array}\right.
\label{physic-spec}
\eeqa
It contains a triplet of massless modes for $l=1$, corresponding
to the Goldstone modes of the broken rotational 
symmetry $SO(4)\to SO(3)$. 
All higher modes are massive, indicating the harmonic stability of the 
D2-brane.

Branes can be viewed also as a bound state of D0-branes. For small WZW branes
($\la\ll k$) the physics of the
system is provided by the ordinary matrix model \cite{myers}. 
In \cite{PS1,PS2} another matrix model of WZW branes
based on quantum symmetries was
proposed. It was shown there that the model properly describes the static
properties of {\em all} branes. For G=SU(2) the  
variables of the model consist of 4 matrices $M^\mu$ 
(for $\mu=1,0,-1, 4,$ and $i,j,k=1,0,-1$),
subject to the relations
 \beqa
F^l_L(M)&\equiv&i( q M^4 M^l-q^{-1}M^l M^4)-\ep^l_{ij} M^i M^j=0 \nn\\
F^l_R(M)&\equiv&i(-q^{-1} M^4 M^l+ q M^l M^4)-\ep^l_{ij} M^i M^j=0.
\label{FLR}
\eeqa 
Here $\ep^k_{ij}$ is the $q$-deformed epsilon-tensor, which 
is recalled in Appendix \ref{basicUq} 
along with the other $q$-deformed objects needed. 
These relations were obtained  in \cite{PS1} by requiring 
invariance under  a ``twisted'' quantum symmetry
$U_q(so(4))_\cF$.
The $M^\mu$ should be thought of as quantizations of 
the coordinate functions $x^\mu$ of the embedding space
$\R^4$. They can also be written as $2\times 2$ matrices 
in terms of the four ($q$-) Pauli matrices 
$\si_\mu=(1,-i q^{-1} \si_i)$
\beq\label{m-mat}
M=M^\mu \si_\mu\equiv M^4-iq^{-1}M^D.
\eeq
where $\tr_q(M^D)=0${} \footnote{note that $M^D$
coincides with the Dirac operators on the quantum sphere
constructed by Bibikov and Kulish \cite{BiKu}} 
\footnote{lower-case $\tr_q$ means  q-trace for the spin $1/2$
  representation, see (\ref{q-trace})}
In this $2\times 2$ notation, the relations 
\refeq{FLR} take the form
$R_{21} M_2 R_{12} M_1 = M_1 R_{21} M_2 R_{12}$ 
(in short-hand
notation), which is known as reflection equation (RE).
Here $R_{12}$ is the so-called R-matrix of $U_q(su(2))$ in 
the fundamental representation. 
There is yet another way of writing 
these relations, which will be used extensively throughout this
paper. We also introduce $\tilde\si_\mu=(1,i q\ \si_i)$ and
\beq\label{mt-mat}
\tilde M=M^\mu \tilde\si_\mu\equiv M^4+iqM^D.
\eeq
We can then split $M\Mt$ and $\Mt M$ into trace and traceless parts
$F_{L,R}=F_{L,R}^l\;\si_l, \ l=1,2,3$, 
\beq\label{split}
M\Mt=\det_q(M)+F_L(M),\quad\Mt M=\det_q(M)+F_R(M).
\eeq
Notice that for $\mg=su(2)$, 
$\det_q(M)= \tr_q(M\Mt)/[2]=\tr_q(\Mt M)/[2]=(M^4)^2 + M^i M^j g_{ij}$.
Then \refeq{FLR} are equivalent to
\beq\label{FLR-M}
F_L(M)=F_R(M)=0.
\eeq

The algebra defined by the relations 
\refeq{FLR} resp. \refeq{FLR-M} has two 
central elements, which provide the basis of the interpretation
in terms of $D$-branes. 
The first one is the quantum determinant
which is
invariant under the full $U_q(so(4))_\cF$ symmetry algebra.
We shall use this to impose the constraint
\beq\label{constr-m}
\det_q(M) = r^2
\eeq
which in a sense defines a 3-sphere of  radius $r$. For simplicity we shall
take $r=1$ from now on. 
The second central element is $M^4$, 
which is invariant only under the ``vector'' subalgebra
$U_q(su(2))^V$ of $U_q(so(4))_\cF$. 

The irreps of the RE algebra coincide with those of 
$U_q(su(2))$, and are labeled by a spin 
$\la\in \{0,\frac 12, ..., \frac k2\}$.
For a given irrep $\la$, 
each $M^i$ is a $(2\la+1) \times (2\la+1)$ matrix,
and the Casimir $M^4$ takes the value
\beq
M^4_\la=\frac 1{[2]_q}(q^{2\la+1}+q^{-2\la-1})
=\frac{\cos\frac{(2\la+1)\pi}{k+2}}{\cos\frac{\pi}{k+2}}.
\label{M4-eigenvalue}
\eeq
This fits nicely with  \refeq{position}, and allows to 
identify this irrep with the stable brane $D_{2\la}$.
Further support for this identification 
and more details can be found in \cite{PS2}.

\section{Degrees of freedom and symmetries}
\label{sec:degsym}

We imagine that in order to describe excitations of the branes, we need
some kind of 
field theory living on the NC algebras defined above. As we do not have any
general procedure applicable in this case, we shall naively follow certain
guidelines stemming from the standard $D0$-brane matrix model \cite{myers}
and other noncommutative gauge theories. These models
show that it is useful to combine the 
matrix fields $A^\mu$ and the matrix background
$M^\mu$ into a single variable which we shall call $B^\mu$. 
These will be general
matrices subject only to some reality conditions (to be defined later). In
other word, we shall split
\beq
B^\mu=M^\mu+A^\mu
\eeq
where $M^\mu$ is given as in the previous section, 
and $A^\mu$ is arbitrary (but in a
certain sense small if corresponding to excitations).
The four matrices $B^\mu$ can again 
be assembled into a $2\times 2$ matrix 
$B,\ \tilde B$ and we can define $B^D,\ F_{L,R}(B)$ according to (\ref{m-mat}-\ref{split}).
They shall transform as  $B \to \pi(u^L) B \pi(Su^R)$ under $U_q(so(4))_\cF$,
where $\pi$ is the spin $1/2$ \rep of $U_q(su(2))$.\footnote{
One can show that 
$\tilde B$ transforms as $B^{-1}$, and
$$
B\Bt \rightarrow \pi(u^L_1) B \Bt \pi(S(u^L_2)), 
   \quad \Bt B \rightarrow \pi(u^R_1) \Bt B \pi(S(u^R_2))\nn,
$$
which is the basis for writing down invariants.}
In this paper we will impose the constraint 
\beq
\det_q(B) = 1,
\eeq
setting the radius $r=1$ for simplicity.
This important point will be discussed in Section \ref{sec:dyn}.

The simplest nontrivial action is now \cite{PS1}
\beq
S[B] = a_L S_L[B]+a_R S_R[B]=a_L \Tr_q(F_L(B) F_L(B))+a_R
\Tr_q(F_R(B) F_R(B)). 
\label{FF-action}
\eeq
for some constants $a_{L,R}$. 
These actions are by construction invariant under $U_q(so(4))_\cF$.
It is understood here that each  $B^\mu$ is a matrix acting on some 
Hilbert space $\cH$, which can be a spin $\la$
\rep of $U_q(su(2))$ for a fixed brane $D_{2\la}$, 
or a more general Hilbert space
as discussed in Section \ref{sec:multibrane}.
Then $\Tr_q$ denotes the quantum trace (see Appendix \ref{basicUq})
over both $\cH$ 
and over the explicit 2$\times$2 matrices; the internal trace can 
be interpreted as integral.
The background configurations $B=M$ respect $F_L(M) = F_R(M) = 0$,
thus they are solutions with $S[M]=0$.

Several remarks are in order. First, the quantum trace 
over $\cH$
guarantees invariance under the following  action of $u \in U_q(su(2))$
\beq
 B^i_j\; \to\;  Su_1\; B^i_j\; u_2, 
\label{cov}
\eeq
where $u_{1,2} \in U_q(su(2))$ 
acts on $\cH$ in the appropriate representations.
For the solutions $B = M$ this transformation is equivalent \cite{PS2}
to the ``vector'' rotations 
$B \to \pi(u_1) B \pi(Su_2)$ for $u \in U_q(su(2))^V \subset U_q(so(4))_\cF$,
hence the action must be 
invariant under \refeq{cov}. Taking 
the classical trace over  $\cH$
instead of the quantum trace would  violate this invariance.

Second, even though 
the action is constructed in terms of invariants of
the quantum group $U_q(so(4))_\cF$,
the precise meaning of a quantum group symmetry in field theory is 
far from trivial. The point is that $S[B]$ is an {\em invariant  expression}
in the sense that it is invariant
if $u \in U_q(so(4))_\cF$ acts on each term using
the coproduct, $u\trr S[B] = S[B]\; \epsilon(u)$ 
(here $\epsilon(u)$ is the counit). 
However the coproduct is nontrivial, hence it is {\em not}
an invariant functional of the matrices: 
$S[u\trr B] \neq S[B]\;\epsilon(u)$.
This will imply 
e.g. that we will not see certain zero modes later.
It probably means that one should  treat the
$B^\mu$'s as some kind of ``dressed'' quantum fields rather than
classical matrices\footnote{Using a covariant differential calculus 
rather than  
``component fields'' $B^\mu$ would not change this issue: the $B^\mu$
can be understood as components of a one-form w.r.t. a frame, 
cp. \cite{qFSI}}. One way of dealing with this problem has been 
proposed in \cite{qFSII} in a simpler context. We will ignore this issue
in the present paper, and treat $S[B]$ in a naive way.  
This should be seen as 
a first step towards a fully consistent treatment.

\paragraph{Star structure.}

In order to have a meaningful theory of branes on $SU(2)$,
we must specify the appropriate reality 
constraints for the gauge fields $B$. 
This should be done in a way which is consistent with the solutions
$B = M$, and compatible with the quantum group symmetries. 
Hence we cannot simply 
impose $B^\dagger = B^{-1}$ or $B^\dagger = B$, 
since $M$ does not satisfy these constraints. 
We define here a different conjugation (star structure) 
respecting $\star^2=1,\ 
\star(XY)=\star Y\star X$, by
\beq
\star(X) := \pi(\wh) \wh X^\dagger \wh^{-1}\pi(\wh^{-1}).
\label{Xstar}
\eeq
Here $X^\dagger = (X^*)^T$, i.e. $(X^i_j)^\dagger = (X^j_i)^*$,
and $x^*$ is the usual adjoint of an operator $x$ acting on a
Hilbert space.
The element $\wh \in U_q(su(2))$ is the ``universal Weyl element'' 
\cite{kirill-resh}
of \uqt, which acts on representations like 
a (q-) mirror reflection of the weights.

It turns out (see appendix) that $\star M=\Mt$,
which\footnote{Also $\star(M) = M^{-1}$.}  implies 
$\star(M_D) = M_D, \; \star(M^4) = M^4$.
In component form, this star becomes 
$\star(M^i) = - M^i$
where now $\star(M^i) =  \wh (M^i)^* \wh^{-1}$. 
We therefore impose the same condition globally on the 
dynamical degrees of freedom:
\beq
\star(B) =  \tilde B,
\label{B-reality}
\eeq
which implies $\star(B_D) = B_D, \; \star(B^4) = B^4$.
It is shown in the appendix that this reality condition is 
preserved under ``vector
rotations'' $B \to \pi(u_1) B \pi(Su_2)$ for $u \in U_q(su(2))^V$, 
as well as under \refeq{cov}.
One can now check 
that the determinant $det_q(B) \propto tr_q(B \Bt)$ is real, 
and so are the 
actions $\Tr_q(FF) = \Tr_q(F \star F)$ using \refeq{trace-reality}
and 
\beq
\star(F) = F.
\label{F-real}
\eeq 
More details are given
in the appendix.

\subsection{Static multi-brane configurations}
\label{sec:multibrane}

We also want to consider configurations of several branes and 
their excitations. Recall that a single static brane 
corresponds to an irreducible \rep  of RE, hence also of $U_q(su(2))$.
Reducible \rep $M=\oplus_\la M_\la$ therefore
describe several branes, i.e.
\beq\label{msystem}
M = \left(\begin{array}{cccc}
        M_{\la_1 } &0 & ... & 0\\
        0 & M_{\la_2 } & ...& 0 \\
                       & ...           &   & \\
       0 & .... & 0 & M_{\la_{N}} 
\end{array} \right).
\eeq
It is then natural to 
decompose 
the field $B$ which describes the
interaction and excitations of such a system in the form
\beq\label{bsystem}
B = \left(\begin{array}{cccc}
        B_{\la_1 } & B_{\la_1 \la_2} & ... & B_{\la_1 \la_N} \\
        B_{\la_2 \la_1} & B_{\la_2 } & ...&  B_{\la_2 \la_N} \\
                       & ...           &   & \\
       B_{\la_N \la_1} & .... & B_{\la_N \la_{N-1}} & B_{\la_{N}} 
\end{array} \right),
\eeq
reflecting that there are $N$ distinct ``positions'' for the branes.
Notice that $B_{\la \g}\in (\la\otimes \frac12)\otimes(\g\otimes \frac12)^*$
while 
$B_\la\in (\la\otimes \frac12)\otimes(\la\otimes \frac12)^*$.
Requiring that single-brane configurations are consistent, 
we should take the reducible representation
of $\wh$, so that 
\beq
\wh = \left(\begin{array}{ccc} \wh_{\la_1} & 0 & 0  \\
                              & ... &  \\
                             0 & 0 & \wh_{\la_N} \end{array}\right).
\eeq
Also the quantum trace should then be defined block-wise\footnote{In 
principle, there is an equivalence map $D_{2\la} \to D_{2(k/2-\la)}$,
defining some automorphism $\a$. One could then also impose
$\star(B_D^i) = - \a(B_D^i)$ etc.}.
We therefore extend the star operation on $B$ as
\beq
 \star(B^4_{\la\g})=B^4_{\g\la} ,
\quad \quad \star(B^D_{\la\g})=B^D_{\g\la} 
\eeq
hence
$
B_{\la \g}^i= -\star(B^i)_{\g \la} =- \wh_{\g}\;
(B_{\g\la}^i)^* \; \wh_{\la}^{-1} 
$
where $*$ means the usual adjoint of an operator.

(This is not the only possibility: one could also take a \rep 
of a given size $N$, which might 
correspond to the number of $D0$-branes.).

\section{Brane dynamics}
\label{sec:dyn}

We now turn to the dynamics of these branes.
Hence we consider fluctuations of the form
\beqa\label{B-expand}
B=(M^4 + A^4) - iq^{-1}(M^D+A^D), \quad \Bt =(M^4+A^4) + iq(M^D+A^D).
\eeqa
For the off-diagonal
entries  in \refeq{bsystem} the background is $M_{\la_1\la_2}=0$.
All the fluctuations around the background are in
$F_L$, due to the determinant constraint.
Recalling that $F_L(M)=0$, they are
\beq\label{f-lin}
F_L(B)=i(qA^4M^D-q^{-1}M^D A^4)+Q\trr A^D-(Q\trr A^D)^{(4)}+O(A^2)
\eeq
where 
\beqa
(Q\trr A^D)^{(4)}&\equiv&\tr_q(Q\trr A^D)/[2]\\
\non
Q\trr A^D&\equiv& i(qM^4A^D-q^{-1}A^DM^4)+M^DA^D+A^DM^D.
\eeqa
We have split $F_L$ into several pieces
in \refeq{f-lin} according to their importance for the
calculations we are going to perform. We shall elaborate on this in the course 
of the paper. Here we just note that the first term in
\refeq{f-lin}
can be neglected, and the most important part will be 
$Q\trr A^D-(Q\trr A^D)^{(4)}$.
We can get similar formulae for $F_R$.

We decided to set $\det_q(B)=r^2=1$ in this paper. 
There are several reasons for 
this. First of all the 
constraint reduces the number of the excitations by one. The
radius $r$ is in principle determined by the background, and can 
easily be reinserted if desired. 
To first order in $A$'s, the constraint $\det_q(B)=1$ implies
\beq\label{am-const}
M^4A^4+A^4M^4+(Q\trr A^D)^{(4)}=0,
\eeq
which 
determines $A^4$ in terms of $A^D$. Furthermore, 
other calculations \cite{jbds} show  that there are 
massless excitation modes corresponding to non-trivial rotations of
the spherical D2-brane inside $S^3 = SU(2)$. 
Such a rotation (even an infinitesimal one,
since the radius of $SU(2)$ is finite here) 
changes the $x^4$ resp. $B^4$ coordinate, 
thus we need $A^4$ to be determined from $\det_q(B)=1$. In 
this paper we shall use only the linear approximation 
\refeq{am-const}, leaving a complete implementation of the
constraint $\det_q(B)=1$ open. 
Notice also that \refeq{am-const} is invariant 
under gauge transformation while $\det_q(B)$ is not.

Let us summarize our setup: 
we assume a background respecting
$F_L(M)=F_R(M)=0$, and impose $\det_q(B)=1$.  From
\refeq{FF-action} and the expansion \refeq{B-expand} the 
action for excitations  becomes
$a_L \Tr_qF_L(B)^2+a_R
\Tr_qF_R(B)^2$ 
with $F_L(B)$ given by \refeq{f-lin},
and an analogous expression for $F_R$.  Since both terms will give
basically the same results, we shall work mostly with $\Tr_q(F_L(B)^2)$
and set $a_L=1$.  Thus we will study the action
\beq
S_L= Tr_q(F_L(B)^2)
\eeq

\subsection{Spectrum of single-brane fluctuations.}
\label{single}

In this subsection we shall determine the 
excitation spectrum on a single brane.
We will consider the action as an ordinary functional of the 
fluctuation matrices $A^\mu$, 
and use a harmonic expansion for the matrix
entries. While the intermediate steps and formulae
in the calculation are adapted to this new context, 
we shall see that the expansion has very similar properties 
to ordinary harmonic analysis.

The result of the calculations will be the masses of brane
excitations. For low harmonics, they
will differ from the known results \cite{jbds}. 
The most important point is that there
there will not be any massless modes corresponding to rigid rotations
of the brane. 
We shall comment on this point in the end of this subsection.
Apart from the  zero modes corresponding to gauge invariance
all masses are positive, hence the branes are stable in our model.

In order to facilitate the 
calculations we shall work here only in the large $k$ expansion,
which is good enough to exhibit the main properties of the spectrum.
In this limit,
all q-tensors appearing in \refeq{FLR},\refeq{FF-action} 
can be replaced by their undeformed counterparts;
thus  $g_{ij}\to\d_{ij}$ and $\eps^l_{ij}\to\eps^c_{ijl}$ 
where $\eps^c_{ijl}$ 
is the ordinary antisymmetric tensor. Then
$F_L=F_R=0$ are solved by 
$M^l= -i(q-q^{-1}) M^4 x^l$ where $[x^i, x^j]=i\eps^c_{ijl} x^l$, 
i.e.  the $x$'s satisfy the ordinary $su(2)$ algebra 
in the spin $\la$ representation (corresponding to the brane).
Assuming that the spin $j$ of the fields
respects $j\ll k$ we can compare the contributions of the various terms,
counting powers of $k$.
Then we get
$Q\trr A^D\sim A^D\ M^4/k$ and $A^4\sim A^D/k$, thus
\beq
i(qA^4M^D-q^{-1}M^D A^4)\sim \frac{M^4}{k^3}A^D
\eeq
and we can neglect this term compared to $Q\trr A^D$. Moreover
$S_L=S_R$ in the case of a single brane 
$D_{2\la}$, 
since $M^4$ is central. Thus we shall discuss only $S_L$.

Therefore the  mass matrix for excitations we are going to consider 
for a brane $D_{2\la}$ is
\beq
S_L = \Tr_q(F_L F_L)=\Tr_q[(Q\trr A^D) (Q\trr A^D) -(Q\trr
A^D)^{(4)}(Q\trr A^D)^{(4)}].
\label{FF-action-QQ}
\eeq
The star was omitted here, since $\star(Q\trr A^D) =Q\trr A^D$ 
due to the reality constraints.
Using the fact that the quantum trace is cyclic\footnote{this follows 
from the explicit realization 
$M^D = (\pi_\g\tens \pi_{\frac 12})(\cR_{21} \cR_{12})$ where $\cR$ is the 
universal $R$ matrix of $U_q(su(2))$, which commutes with $q^H\tens
q^H$} 
with respect to  $M^D$, 
this can be rewritten as
\beqa
S_L =\Tr_q[ A^D (Q^2\trr A^D) - (Q\trr A^D)^{(4)} (Q\trr A^D)^{(4)}]
\label{FF-action-Q}
\eeqa
where
\beq
Q^2 \trr A^D = (\a^2 + 2\b) A^D - \a (M^D A^D + A^D M^D) + 2 M^D A^D M^D
\eeq
using the characteristic equation \refeq{chareq} on the brane,
\beqa
(M^D)^2 &=& \a M^D + \b, \label{char-eq}\\
\a &=& - h  M^4, \; \b = 1-(M^4)^2, \nn\\
h &=& i(q-q^{-1}).
\eeqa

The action for the quadratic fluctuations
has the following infinitesimal gauge invariance
\beq\label{gauge}
\d A^D = i(M^D f - f M^D)
\eeq 
for any $f\in Mat(2\la+1)$,
i.e. $(Q\trr \d A^D) =0$.
This gauge invariance is not a standard consequence of
symmetries of the action but rather its equations of motion
$F_L(M)=0$ and the fact that the action is a functional of $F_L$.  Notice that
after \refeq{gauge} we get (for infinitesimal $f$) 
$B=M+A\to e^{-if}M e^{if}+A$ thus  
terms linear in $A$ in the expansion \refeq{f-lin} of $F_L(B)$ are not changed
because $e^{if}M e^{-if}$ also solves equations of motion. The argument 
works for any functional depending solely on $F_L$ or $F_R$.

\subsubsection{Harmonic expansion}

It is very important to choose a convenient basis for the gauge fields
$A$. 
The latter are elements of $Mat(2\la+1)\otimes Mat(2) $ 
and can be represented in the
following manner: $A^D=A_{(1)}+A_{(2)} + A_{(g)}$ where
\beqa\label{basis}
A_{(1)} &=& M^D f + f M^D, \\
A_{(2)} &=&  M^D f' M^D - (M^D f' M^D)^{(4)} \nn \\
A_{(g)} &=&  i(M^D f'' - f'' M^D)  \nn
\eeqa
where $f,\ f'$ are ``scalar fields'' with values in $Mat(2\la+1)$. 
$A^4$ is then determined by \refeq{am-const}.
We can discard the gauge degrees of freedom $A_{(g)}$ as discussed
above. 
Moreover the fields $A_{(i)}$ 
satisfy the reality constraints \refeq{B-reality}
provided $\star(f) = \wh f^* \wh^{-1} = f$. 

We now expand $f\in Mat(2\la+1)$ in terms of
harmonics, i.e. eigenvalues of a suitable Laplacian. 
The following definition turns out to be useful:
\beq
\Delta f = [\la_i, f] \la^i = -\frac 12 [\la_i,[\la^i,f]] 
\eeq
where $M^D = \la_i \sigma^i$, noting that $\la_i \la^i$ is central.
From the algebra relations \refeq{char-eq},
it follows that $J_i = \frac{\la_i}{\a}$ satisfies 
$i\eps J J = J$. 
Furthermore, $\star(\Delta f) =  \Delta\star f$.
In order to simplify the calculations below, we shall work in the
leading order in $1/k$ 
for the rest of this subsection.
Then the eigenvalues of $\Delta$ are
to lowest order given by
\beq
\Delta f_{jm} = -\frac 12 \a^2\; j(j+1)\; f_{jm} \quad (\; +\;o(\frac 1k)).
\eeq
Moreover we can normalize the harmonics in the standard
way\footnote{because
the Laplacian is self-adjoint w.r.t. the inner product $\Tr_q(f \star g)$}
$\Tr_q(f_{jm}f_{j'm'})=\d_{jj'}\d_{mm'}$. 
Any $f\in Mat(2\la+1)$ can now be represented as
\beq
f=\sum_{jm} a_{jm}f_{jm}.
\eeq
With this we get 
\beq
A_{(2)} =  M^D f' M^D - (\b + \Delta)f' 
\eeq
where $\Delta$ becomes a number if $f'$ is a harmonic.

Notice that contribution to \refeq{FF-action-Q} of different
harmonics are orthogonal 
to each other.  
This can be seen as follows:
after expanding $A^D$ in harmonics ($f,\ f'$),
any term in \refeq{FF-action-Q} is of the form (using cyclicity for $M^D$)
\beq
\Tr_q((M^D)^m f (M^D)^n f').
\eeq
Using the characteristic equation \refeq{char-eq},  
one can furthermore simplify these to
\beq
\Tr_q(f f'),\ \Tr_q( f M^D f'),\ 
\Tr_q(M^D f  f'),\ \Tr_q(M^D f M^D f').
\eeq
The second and the third one vanish due to $\Tr_q$, while the last one is
proportional to $\Tr_q(f f')$ since $\Tr_q(M^D f M^D f')=\Tr_q((M^D f
M^D)^{(4)} f')=(\b + \Delta) \Tr_q(f f')$.
This vanishes if $f$ and $f'$ 
are different harmonics.

To summarize, we can restrict ourselves to quadratic fluctuations of the form
\beq\label{expa}
A^D=a_1 A_{(1)}+a_2 A_{(2)}
\eeq 
where $a_i\in \R$ and $f=f'$, $\Delta f = -\frac 12 \a^2\; j(j+1)\; f$.
Then the action \refeq{FF-action-QQ} becomes
\beq\label{action-2}
S = a_i a_j\ \Tr_q(Q\trr A_{(i)}^D\; Q\trr A_{(j)}^D 
- (Q\trr A^D_{(i)})^{(4)} (Q\trr A^D_{(j)})^{(4)}).
\eeq
The calculation now proceeds using the same tricks as above.
It is convenient to introduce
the matrix of normalizations\footnote{
Notice that $G_{ij}$ is not singular except one case $j=0$. In this case
$A_{(1)}$ and $A_{(2)}$ are dependent thus we must remove one them.} 
for the modes $A^D_{(i)}$,
\beqa
G_{ij}\equiv \Tr_q(A^D_{(i)} A^D_{(j)})= \left(\begin{array}{cc} 4\b +2\Delta,
    & 2 \a(\b+\Delta) \\ 
                      2 \a(\b+\Delta), & \a^2 \b +(\a^2 -2\b)\Delta -\Delta^2
\end{array}\right).
\eeqa
Here $\Delta$ will always stand for its eigenvalue on $f$ resp. $f'$. 
We furthermore need
\beqa
(Q \trr A_{(1)})^{(4)} &=& (4\b+2\Delta) f, \nn\\
(Q \trr A_{(2)})^{(4)} &=& 2\a(\b+\Delta) f', \nn\\
(Q^2 \trr A_{(1)})^D &=& 4 \b A_{(1)} + 2 \a A_{(2)} \nn\\
(Q^2 \trr A_{(2)})^D &=&  (\a^2- 2 \Delta) A_{(2)} 
  + (2 \a\b+ \a\Delta) A_{(1)}.
\eeqa
Introducing
\beqa
\tilde Q = \left(\begin{array}{cc} 4\b, & 2\a\b +\a\Delta \\
                               2\a, & \a^2 -2\Delta \end{array}\right)
\eeqa
the action \refeq{action-2} becomes 
\beqa\label{mass-l}
a_i a_j \left[ G \tilde Q  - \left(\begin{array}{c} 4\b +2\Delta\\2\a(\b+\Delta)
  \end{array}\right) \left(\begin{array}{cc} 4\b +2\Delta, & 2\a(\b+\Delta)
  \end{array}\right)\right]_{ij}    =a^T\  G T\ a
\eeqa
where 
\beq
T = \left(\begin{array}{cc} -2\Delta, & -\a\Delta \\
         2\a, & \a^2 -2\Delta\end{array}\right). 
\eeq
The matrix $T$ has eigenvalues 
\beqa
t_1 &=& \a^2\; l^2, \qquad\quad  l = 1, 2, ..., 2\la \nn\\
t_2 &=& \a^2\; (l+1)^2,  \quad l = 0, 1, ..., 2\la 
\eeqa
which are the mass spectrum of gauge
fields\footnote{To see this, assume that we use an 
orthonormal basis  $A^o_{(i)}$ instead of
\refeq{basis} and the expansion \refeq{expa} reads $A^D=b_1 A^o_{(1)}+b_2
A^o_{(2)}$. 
Then we can write $G=g^T g$ and $b_i=g_{ij} a_j$. Thus
\refeq{mass-l} is $a^T\  G T\ a=b^T\ g\ T\ g^{-1} b$, and the 
eigenvalues of $ g\ T\ g^{-1}$ and $T$ 
are the same yielding the masses.}.
Each value has the usual $2l+1$ degeneracy, and
$l=0$ must be excluded from the first series because 
$A_{(1)}$ and $A_{(2)}$ coincide in that case.
In particular, all masses are positive, reflecting the stability
of the branes. 

This spectrum  should be compared with \refeq{physic-spec}. 
We see that 
the result is very close (including the correct scaling dependence on $k$),
but for small $l$ it is not the same. 
Most notably we are missing the triplet of zero-modes in 
\refeq{physic-spec}, which correspond to the nontrivial rotation 
of the 2-branes in $S^3$. 
The lack of rotational zero-modes indicates that our action is 
{\em not} invariant under $SO(4)$. This may seem strange, 
because the action was constructed to be 
invariant under a quantum version of that symmetry, 
$U_q(so(4))_\cF$. This was successful in the sense that it leads 
quite directly to equations of motion (RE) which have the 
correct brane solutions. We recall however the discussion 
in Section \ref{sec:degsym} that this symmetry is formal -- in our
naive treatment as ordinary matrix model -- and
not a symmetry of the action functional: 
$S[u\trr B] \neq S[B]\;\varepsilon(u)$.
This indicates that a fully consistent way of implementing 
this quantum symmetry is still to be found.

\subsection{Strings stretched between branes}
\label{inter}

We now discuss several branes, and calculate masses of states 
which 
mediate interactions between them. 
We must therefore consider reducible 
\reps of $M$, and include
the ``off-diagonal'' sectors $B_{\g \la}$
which connect different branes. 

Assume that there are 2 branes
$D_{2\g}$ and  $D_{2\la}$ present. We want to calculate the
lowest energy (mass) for a ``string'' connecting these 2 branes.
This scenario is described by the matrix
\beq\label{2-brane}
B = \left(\begin{array}{cc}
         M_{\g}  & A_{\g\la}   \\
         A_{\la\g}  &  M_{\la}  \end{array} \right).
\eeq
One could now in principle do a similar analysis as for the 
single-brane modes. To simplify the calculation, 
we shall compute here the ground state energy (mass) only,
using a somewhat different approach.

To proceed, it is useful to introduce a basis of eigenvectors 
of the ``Dirac operators'' $M^D$. We consider here $M^D$
as acting on 
$(\g \otimes \frac12)$  from the left 
and on $(\la\otimes\frac12)^*$ from the right. 
The star refers here to the way $U_q(su(2))$ acts, and will be explained below.
The left eigenvalues of $M^D$ are \cite{BiKu}
\beq\label{dir}
M^D\ |\g\a,m\ran =c^\a_\g\ |\g\a,m\ran \quad \in (\g\otimes \frac 12) 
\eeq 
where $\a = \pm \frac 12$, 
$m \in \{-\g, ..., \g\}$,
$c^{\pm 1/2}_\g=\pm h\ [\g+(1\mp 1)\frac12]_{q^2}$
and $h = i(q-q^{-1})$.
This follows e.g. from the characteristic equation
\refeq{char-eq} of $M^D$, 
which implies that there are also right eigenvectors
with the same eigenvalues:
\beq\label{dir2}
\lan\la\b,n| \ M^D =c^\b_\la\ \lan\la\b,n|.
\eeq
To understand the transformation properties of these eigenvectors, 
we recall the basic fact that  $M^D$ commutes with the coproduct 
$\Delta(u) = u_1 \tens u_2$ of $u \in U_q(su(2))$:
\beqa
(u_1 \tens u_2) M^D &=& M^D (u_1 \tens u_2), \label{comm1}\\
M^D (Su_2 \tens Su_1) &=& (Su_2 \tens Su_1) M^D \label{comm2}
\eeqa
where the appropriate \reps ($\pi_\g, \pi_{\frac 12}$) are 
understood\footnote{this follows from the explicit realization 
$M^D = (\pi_\g\tens \pi_{\frac 12})(\cR_{21} \cR_{12})$}.
This implies that the left eigenvectors $|\g\a,m\ran \in (\g\otimes\frac12)$ 
transform under the following action of $U_q(su(2))$
(below  $\a,\b = \pm \frac 12$):
\beq
u \trr (|p\rangle \tens |\a\rangle)
 := u_1|p\rangle \otimes u_2|\a\rangle
  \quad \in (\g\otimes \frac 12) 
\eeq
This justifies the notation in \refeq{dir} using the quantum number $m$.
Similarly, \refeq{comm2} implies that 
the right eigenvectors $\lan\la\b,n| \in (\la\otimes \frac 12)^*$ 
transform under the following ``dual'' action of $U_q(su(2))$:
\beq
u \trr (\lan l| \tens \lan \b|)
 := \lan l| Su_2 \otimes \lan\b| Su_1
  \quad \in (\la\otimes \frac 12)^*. 
\eeq
Therefore the matrices
\beq
\xi_{\g\a m; \la\b n} := |\g\a,m\ran\lan\la\b,n|
\eeq
transform as
\beq
u \trr \xi_{\g\a m; \la\b n} = (u_1 \tens u_2) 
\xi_{\g\a m; \la\b n} (Su_4 \tens Su_3) \;\; 
\in (\g\otimes \frac 12)\otimes (\la\otimes\frac 12)^*,
\eeq
and one can decompose them accordingly into irreps of $U_q(su(2))$.
Furthermore, $\star(M^D) = M^D$ implies that 
$\star \xi_{\g\a m; \la\b n}$ 
is also ``eigenmatrix'' of $M^D$ with flipped left and right eigenvalues, 
and one can check (see appendix) that
\beq
\star(u \trr \xi_{\g\a m; \la\b n}) = u \trr \star\xi_{\g\a m; \la\b n}
\eeq
provided $u$ is in the following  ``real'' sector
of the rotation algebra (see also \refeq{real-group})
\beq
\cG^V  := \{u = \theta u^* = \wh (Su^*) \wh^{-1} \}.
\eeq
This means that the star $\star$ restricts consistently
to irreps of $U_q(su(2))$.

Armed with these tools, we return
to the gauge fields $A_{\g\la}$ which a
priori are arbitrary matrix valued fields in 
$(\g\otimes \frac 12)\otimes (\la\otimes\frac 12)^*$. 
We can therefore expand them in the basis of eigenvectors
of the  Dirac operator \refeq{dir}: 
\beq
A_{\g\la}=\sum a(\g\a,m;\la\b,n)\; \xi_{\g\a m; \la\b n}
\label{Amn-exp}
\eeq
for arbitrary $a(\g\a,m;\la\b,n)$. Note that in general,
the matrices $\xi_{\g\a m; \la\b n}$ have non-vanishing 
trace, i.e. in general $A^4 \neq 0$ in the above expansion.
However, we are interested only in the ground states 
(which we assume to be the minimal spin states) here.
Assume furthermore that $\g\geq\la+1$. Then the 
spin of the ground state  
is $(\g-\la-1)$. This implies that there is a unique such multiplet,
and it must be in $1 \otimes\g\otimes\la\subset(\g\tens \frac 12)\otimes 
 (\la\otimes \frac 12)^*= ((\g-\frac12) \otimes (\la+\frac12))\oplus
 ...=(\g-\la-1)\oplus ...$ where dots denote the higher spin states.
Therefore the ground state has the form
\beq
A_{\g\la}^D = A_{\g\la} = \sum a(\g-\frac 12,m;\la+\frac 12,n)\;
|\g-\frac 12,m\ran\lan\la+\frac 12,n|
\label{Amn-ground-exp}
\eeq
with $A^4=0$, since the singlet in the spinor part does not
enter\footnote{this argument requires the detailed discussion
of the transformation properties given above}.

We can now calculate the eigenstates of the mass matrix for 
the ground states of the 
off-diagonal excitations in \refeq{2-brane}.
We shall concentrate on $S_L$, and one can check that $S_R$ gives
the same result. 

To get a real gauge field one must also include the conjugate term 
$A_{\la\g}$. Hence we should consider
\beqa\label{massm}
\Tr_q[(Q\trr A^D)_{\g\la}*(Q\trr A^D)_{\g\la}]+(\la\leftrightarrow\g)
\eeqa 
for fields of the form \refeq{Amn-ground-exp}.
Since for the ground states $\xi$ is traceless as discussed above,
we can ignore the terms $(Q\trr A^D)^{(4)}$ 
in the action \refeq{FF-action-QQ}. 
It is easy to see that 
\beq
Q\trr\xi_{\g\a m; \la\b n}=
[i(qM^4_\g-q^{-1}M^4_\la)+ h(c^\a_\g+c^\b_\la)]\xi_{\g\a m; \la\b n}
\eeq
where $\a,\b=\pm\frac12$ according to signs appearing in \refeq{dir}.
$Q$ is not hermitian so the eigenvalue is not real.  
We also have
\beq
\star(Q\trr\ \xi_{\g\a m; \la\b n})=
[i(qM^4_\g-q^{-1}M^4_\la)+h(c^\a_\g+c^\b_\la)]^*\ 
\star\xi_{\g\a m; \la\b n}.
\eeq 
Thus the mass matrix for these excitations $A^D=\xi$ has the
following form
\beq
Q\trr A^D\star(Q\trr A^D)=m^2\ \xi\star\xi
\eeq
with eigenvalues
\beq  
m^2=|i(qM^4_\g-q^{-1}M^4_\la)+h(c^\a_\g+c^\b_\la)|^2
\eeq
This is indeed large for $\a = \b$ and small for $\a = -\b$,
as expected. Hence for 
$A^D = \xi = |\g-\frac 12,m\ran\lan\la+\frac 12,n|^*$, 
the mass squared is
\beq\label{mass-inter}
m^2=|i(qM^4_\g-q^{-1}M^4_\la)+h([\la]_{q^2}-[\g+1]_{q^2})|^2=
4\sin^2\frac{(\g-\la)\pi}{k+2}
\eeq
using \refeq{M4-eigenvalue}.
For large $k$ and $(\g-\la)\ll k$ \footnote{recall that
  $\g\in[0,k/2]$.}, 
this is
\beq
m^2\approx 4 (\frac{(\g-\la)\pi}{k+2})^2,
\eeq
which is indeed the  (arc) length squared of the string
stretched between branes, up to corrections of order $\frac 1{k^2}$.

It is quite remarkable that this simple matrix model 
correctly reproduces the curvature effects of the underlying space $S^3$.
This nicely supports the basic idea of our approach, which is 
to describe the dynamics of $D0$ branes in terms of a matrix
model based on certain ``quantum'' symmetries. 
It would of course be very interesting to check whether the deviation 
of the ground state energy \refeq{mass-inter} 
from the arc-length is in agreement with string theory, 
perhaps due to the $B$ field which is not closed.

The attentive reader might have noticed that the calculation presented
here is not valid for $\g = \la + \frac 12$. We expect that the result
would be the same, which could e.g. be verified using an approach
similar as in Section \ref{single}.

\section{Discussion}

In this paper we analyzed the dynamics of the matrix model
for $D$-branes on $SU(2)$ which was 
proposed in \cite{PS1,PS2}, based on noncommutative algebras
related to quantum groups. This was motivated by 
the algebra of boundary operators which involves the 
$6j$-symbol of $U_q(su(2))$, and exhibits a 
particular truncation pattern related to that quantum group \cite{ars2}. 
Our matrix model was designed to work for finite $k$
as opposed to the ones found in \cite{ars2}.

The main result of this paper is to show that this model also gives a 
reasonable description for the dynamics of noncommutative $D$-branes. 
In particular, the branes turn out to be stable in our model. 
In doing so we tried 
to keep things simple, and stayed as close 
as possible to the usual treatment of matrix models. This includes
ignoring some inconsistencies, 
in particular in the context of the quantum group
symmetries.  The price to pay is that we apparently do not get 
everything right, for example our model lacks the zero-modes
associated to the global rotations. 
This could probably be cured if 
one would find a consistent implementation of the symmetries 
on a ``quantum'' level. 

Let us summarize the main merits of the proposed 
models, including the results of \cite{PS1,PS2}:
\begin{itemize}
\item the geometric properties (quantized position, radius, ...)
of the branes as embedded in the group manifold are reproduced,
including the symmetry $\la \leftrightarrow k/2 -\la$
\item each brane is a ``fuzzy'' (noncommutative) space
with the correct fusion rules and truncations in the spectrum of harmonics
\item the energy $\dim_q(V_\la)$ of a brane $D_{2\la}$ is naturally
obtained by adding $\Tr_q(1)$ to the action 
\item the ground state energy for strings stretched between different
  branes seems (essentially) correct
\item each brane is stable, with mass spectrum close to the correct
  one.
\end{itemize}
In view of all this, we believe that the models 
are very interesting as they are, being perhaps 
the first matrix models which describe branes on a curved background.

Of course it would be useful to extend our results to other directions. For
$G=SU(N)$, $N>2$ the 
zoo of branes in the WZW model is
much richer than for $G=SU(2)$,
for example 
there are so-called twisted branes \cite{FFFS}. One could also try to
analyze the coset WZW models. There are plenty of results concerning ordinary 
matrix models for these systems \cite{other} which can be used as 
guidelines for the construction of quantum matrix models.

However, we should also emphasize that the model 
as treated here gives an incorrect
mass spectrum of single branes, 
in particular the rotational zero modes are missing, 
which is a serious drawback. 
One should also keep in mind that there is some freedom in defining the
action, in particular related to the constraint. For example, 
instead of imposing $\det_q(B) = const$ one could replace this 
by its linearized form \refeq{am-const},
$M^4A^4+A^4M^4+(Q\trr A^D)^{(4)}=0$,
which is $U_q(so(4))$ invariant. 
Then
$\tr_q(B\Bt)=\tr_q(M\Mt)+\tr_q(A\tilde A)$
is another possible term in the action, which is 
quadratic in $A$. However, this would not change 
the main features of our results.
Of course one could also consider higher-order terms 
of $F$ in the action.
Furthermore, the choice of reality conditions may affect
the physical content of the models. While this is largely
dictated by the reality property \refeq{B-reality} of the solutions
$B = M$, there may be other possibilities how to extend this
to the fluctuations, 
perhaps requiring additional terms in the (real) action.
However the main problem of the present approach seems to be the
lack of a consistent implementation of this quantum group invariance
beyond the formal level. This could be fixed by ``quantizing'' the 
matrices in an appropriate way (cp. \cite{qFSII}), or by some 
kind of symmetry-respecting Seiberg-Witten map.
Finding a formalism which provides that is a challenge for the future,
which seems worthwhile to pursue.

\section*{Acknowledgements}

We would like to thank P. Aschieri,
H. Grosse, B. Jurco, J. Madore, H. Ooguri, Ch. Schweigert,
A. Sitarz, R. Suszek and J. Wess 
for useful discussions. J.P. would also like to thank  the
Ludwig--Maximilians--Universit\"at M\"unchen and MPI f\"ur Gravitationsphysik,
Albert-Einstein-Institut at Potsdam for hospitality, and 
H.S. thanks the  IFT of Warsaw University for hospitality.

\begin{appendix}
\section{Appendices}

\subsection{Basic properties of $U_q(su(2))$}\label{app:defs}
\label{basicUq}

The basic relations of the Hopf algebra $U_q(su(2))$ are 
\beq\label{U-q-rel}
[H, X^{\pm}] = \pm 2 X^{\pm},  \qquad
[X^+, X^-]   =  \frac{q^{H}-q^{-H}}{q-q^{-1}}=  [H]_{q},
\eeq
where the $q$--numbers are defined as 
$[n]_q = \frac {q^n-q^{-n}}{q-q^{-1}}$. 
The action of $U_q(su(2))$ on a tensor product of representations 
is encoded in the coproduct
\beqa 
\Delta(H)       &=& H \otimes 1 + 1 \otimes H \nonumber \\
\Delta(X^{\pm}) &=&  X^{\pm} \otimes q^{-H/2} + q^{H/2}\otimes X^{\pm}.
\label{coproduct-X} 
\eeqa
We use the Sweedler--notation $\Delta(u) = u_1 \otimes u_2$, where a 
summation convention is understood.
The antipode and the counit are given by
\beqa
S(H)    &=& -H, \quad  
S(X^+)  = -q^{-1} X^+, \quad S(X^-)  = -q X^-, \nonumber \\
\epsilon(H) &=& \epsilon(X^{\pm})=0.
\eeqa
The quantum trace of an operator $A$ acting on 
a representation of $U_q(su(2))$ is defined by
\beq
\tr_q(A) = \tr(A \pi(q^{-H}))
\label{q-trace}
\eeq
where $\pi$ denotes the representation. 
It has the important invariance property  
$\tr_q(\pi(u_1) A \pi(Su_2)) =  \tr_q(A)\ep(u)$ for any $u \in U_q(su(2))$.
This is based on the identity
\beq
S^2(u)=q^{-H} u\  q^{H}
\eeq
for $u \in U_q(su(2))$, which is easy to check.

\paragraph{Invariant tensors.}

The  $q$--deformed sigma--matrices are given by
\beqa
\sigma_{-1} = \left(\begin{array}{cc} 0 & q^{\frac12}\sqrt{[2]_q}\\ 
                            0 & 0 \end{array}\right),  \quad
\sigma_{0} = \left(\begin{array}{cc} -q^{-1} & 0 \\ 
                                                0 & q \end{array}\right), \quad
\sigma_{1} = \left(\begin{array}{cc} 0 & 0 \\ 
                           - q^{-\frac12}\sqrt{[2]_q} & 0 \end{array}\right).
\label{sigma-matrices}
\eeqa
They satisfy
\beqa
\sigma_i \sigma_j &=& - \ep_{ij}^k \; \sigma_k + g_{ij} \label{sigma-cliff}\\
\pi(u_1) \sigma_i \pi(S u_2) &=& \sigma_j \pi^j_i(u),
\eeqa
for $u \in U_q(su(2))$, where $\ep_{ij}^k$ is defined below, and
$\pi$ denotes the appropriate representation.

The invariant tensor $g^{ij}$ for the spin 1 representation 
satisfies by definition
\beq
\pi^i_k(u_1)\; \pi^j_l(u_2)\; g^{kl} = \ep(u) g^{ij}
\label{invar-g}
\eeq
for $u \in U_q(su(2))$. It is given by
\beq
g^{1 -1} = -q^{-1}, \;\; g^{0 0} = 1, \;\; g^{-1 1} = -q,
\label{g-explicit}
\eeq
all other components are zero.
Then $g_{ij} = g^{ij}$ satisfies  
$g^{ij} g_{jk} = \d^{i}_{k}$, and 
$g^{ij} g_{ij} = q^2 +1 + q^{-2} = [3]_q$.

The  Clebsch--Gordon coefficients
for $(3) \subset (3) \otimes(3)$, i.e. the $q$--deformed structure 
constants, are given by 
\beq
\begin{array}{ll} 
\ep^{1 0}_{1} =  q^{-1}, & \ep^{0 1}_{1} = -q,  \\
\ep^{0 0}_{0} = -(q-q^{-1}), & \ep^{1 -1}_{0} = 1 = -\ep^{-1 1}_{0}, \\
\ep^{0 -1}_{-1} = q^{-1}, & \ep^{-1 0}_{-1} = -q,
\end{array}
\label{C-ijk}
\eeq
and $\ep_{i j}^{k} := \ep^{i j}_{k}$. They have been normalized such that 
$\sum_{ij} \ep_{ij}^{n} \ep^{ij}_{m} = [2]_{q^2} \d^n_m$.

\subsection{Star structure}

We give some more details on the star structure here.
Recall the  universal Weyl element \cite{kirill-resh}
$w$ for \uqt, which satisfies 
\beqa
\Delta(w) &=& \cR^{-1} w\tens w = w\tens w \cR_{21}^{-1},\label{del-om-left}\\ 
w u w^{-1} &=& \theta S^{-1}(u),  \quad  w^2 = v
\eeqa
where $v$ is a Casimir and $\theta$ the Cartan-Weyl involution.
For $U_q(su(2))$ we have
\beqa
w\ X_\pm w^{-1}=-q^{\pm 1}X_{\mp},\quad w\ H w^{-1}=-H.
\eeqa
Using this 
we define a rescaled $\wh = w c$ where $c$ is a suitable 
Casimir\footnote{this is not essential, but it simplifies things.}, 
such that 
\beq
\wh^2=1,\quad \quad \wh^\dagger=\wh^{-1} = \wh.
\label{wh-property}
\eeq
The $\star$ is then defined as in \refeq{Xstar},
\beq
\star(X) := \pi(\wh) \wh X^\dagger \wh^{-1}\pi(\wh^{-1}) 
\eeq
where $\pi(\wh)$ acts on the external indices of the matrix $X$
and $\wh$ on the ``internal'' space.
It satisfies $\star \star = \mbox{id}$ due to \refeq{wh-property},
and
\beq
\star(\tr_q(X)) = \tr_q(\star(X))
\label{trace-star}
\eeq
since $q^{H} \wh q^{H} = \wh$.
Then for any such matrices $X,Y$, one has 
\beq
\star(\tr_q(X Y)) = \tr_q(\star(XY)) = \tr_q(\star(Y) \star(X)).
\label{trace-reality}
\eeq
Using $\cR^{*\tens *} = \cR_{21}^{-1}$ together with 
\refeq{del-om-left}, it follows that
\beq
\star(M) = M^{-1}.
\eeq
Moreover, $\star(B)= \tilde B$ implies that 
\beq
[2] \star(B^4) =\star\tr_q(B)= \tr_q(\star(B))= \tr_q(\tilde B)=[2] B^4,
\eeq
and together with 
$\star(B) = \star(B^4) +iq \star(B_D) = \tilde B = B^4 +iq B_D$
it follows that 
\beq
\star(B_D) = B_D, \qquad \star(B^4) = B^4
\eeq
To find the star for the components, 
one can  check that 
\beq
\star(\sigma_i) = -\sigma_i
\eeq
for the $q$-Pauli matrices, so that 
\beq
\star(B^4) = B^4, \quad \star(B^i) = - B^i.
\eeq
For $M$, this can also be verified using
the characteristic equation for $M$ in the 
spin $\la$ irrep,
\beq
(M-q^{2\la})(M-q^{-2\la-2}) =0.
\label{chareq}
\eeq

\paragraph{Consistency of reality constraint with transformations.}

We verify that the above reality constraint
is consistent with the algebra of vector rotations 
$B \to \pi(u_1) B \pi(Su_2)$ for 
$u \in U_q(su(2))^V$, 
which is preserved in the presence of a brane.
We determine how $\star(B)$ transforms 
under vector rotations:
\beqa
\star(B) \to &&\pi(\wh)\wh\;(\pi(u_1) B \pi(Su_2))^\dagger 
  \wh^{-1}\pi(\wh^{-1}) = 
  \pi(\wh)\pi(S((u_2)^*)) \wh B^\dagger \wh^{-1}\pi((u_1)^*) \pi(\wh^{-1}) \nn\\
&& = \pi(\wh) \pi(S((u^*)_1)) \pi(\wh^{-1}) \star(B)\pi(\wh) \pi((u^*)_2) \pi(\wh^{-1}) \nn\\
&&   =  \pi(\theta((u^*)_1)) \star(B) \pi(S\theta((u^*)_2)) 
  =  \pi((\theta u^*)_1) \star(B) \pi(S(\theta u^*)_2)\nn\\
&&   = \pi(u_1) \star(B) \pi( Su_2)
\label{hermitian-gaugetansf}
\eeqa
(using $S \theta = \theta S^{-1}$) provided 
\beq
u \in \cG^V  = 
\{u \in U_q(su(2))^V;\;\; u = \theta u^* = \wh (Su^*) \wh^{-1} \}.
\label{real-group}
\eeq
Therefore $\star(B_D) = B_D$ is preserved under rotations 
$u \in \cG^V$, which 
seems to be the appropriate ``real'' rotation group (algebra)
compatible with \refeq{B-reality}.
$\cG^V$ is closed under addition and multiplication. 
A similar calculation shows consistency with the 
transformations \refeq{cov}.

One can in fact show consistency with the full rotation group
$U_q(so(4))_\cF$. This is most easily done 
in terms of coactions of the dual quantum group
with generators $s,t$ \cite{PS2},
where the rotations take the form $B \to s B t^{-1}$.
Then one can show that $\tilde B \to t \tilde B s^{-1}$, 
hence $\star(t) = t^{-1}$ and $\star(s) = s^{-1}$ will guarantee
that $\star(B)$ transforms as $\tilde B$.

\end{appendix}


\newpage

\begin{thebibliography}{99} 


\bibitem{as} A. Yu. Alekseev, V. Schomerus, ``D-branes in the WZW model'',
{\em Phys. Rev.} {\bf D60} (1999) 061901, {\tt hep-th/9812193}.

\bibitem{gawedzki} K.\ Gaw{e}dzki, ``Conformal field theory: a case study'', 
  {\tt hep-th/9904145};\\
V. Schomerus, ``Lectures on Branes in Curved Backgrounds'',
 {\em Class.Quant.Grav.} {\bf 19} (2002) 5781-5847,  {\tt hep-th/0209241}.

\bibitem{jbds} C. Bachas, M. Douglas, C. Schweigert, 
``Flux Stabilization of D-branes'',
{\em JHEP} {\bf 0005} (2000) 048, {\tt hep-th/0003037};\\
J. Pawe{\l}czyk, ``SU(2) WZW D-branes and their non-commutative geometry
from DBI action'',
{\em JHEP} {\bf  08} (2000) 006, {\tt hep-th/0003057};\\
P. Bordalo, S. Ribault, C. Schweigert, 
``Flux stabilization in compact groups'', {\em JHEP} {\bf 0110} (2001) 036,
{\tt hep-th/0108201}.


\bibitem{myers} R. C. Myers, ``Dielectric-Branes'',
{\em JHEP} {\bf 9912} (1999) 022, {\tt hep-th/9910053}; \\
D. Kabat and W. Taylor, ``Spherical membranes in Matrix theory'',
{\em Adv. Theo. Math. Phys.} {\bf 2} (1998) 181, 
{\tt hep-th/9711078};\\ S.-J. Rey, ``Gravitating M(atrix) Q-Balls'',
{\tt hep-th/9711081}


\bibitem{ars2} A. Yu. Alekseev, A. Recknagel, V. Schomerus, 
``Brane Dynamics in Background Fluxes and Non-commutative Geometry''.
{\em JHEP} {\bf 0005} (2000) 010, {\tt hep-th/0003187}.


\bibitem{tw-K}
A.~Kapustin, ``D-branes in a topologically nontrivial {B-field}'', {\em Adv.
  Theor. Math. Phys.} {\bf 4} (2000) 127--154,
{\tt hep-th/9909089};\\
P.~Bouwknegt and V.~Mathai, ``{D-branes}, {B-fields} and twisted {K-theory}'',
  {\em JHEP} {\bf 03} (2000) 007,
{\tt hep-th/0002023}.

\bibitem{mms} J. Maldacena, G. Moore, N. Seiberg, 
``D-Brane Instantons and K-Theory Charges'',
 {\em  JHEP} {\bf 0111} (2001) 062, {\tt hep-th/0108100}.

\bibitem{moore-rev}
G.~Moore,
``K-theory from a physical perspective,''
 {\tt hep-th/0304018}.

\bibitem{PS1} J. Pawe{\l}czyk, H. Steinacker, 
 ``Matrix description of D-branes on 3-spheres'', {\em JHEP} {\bf 0112} (2001)
 018, {\tt hep-th/0107265}. 

\bibitem{PS2} J. Pawe{\l}czyk, H. Steinacker, 
``A quantum algebraic description of D-branes on group manifolds'',
{\em Nucl.Phys.} {\bf B 638} (2002) 433-458,  {\tt hep-th/0203110}.


\bibitem{FFFS} G. Felder, J. Fr\"ohlich, J. Fuchs, C. Schweigert, 
``The geometry of WZW branes'', {\em J.Geom.Phys.} {\bf 34} (2000) 162-190,
{\tt  hep-th/9909030} 

\bibitem{BiKu} P.N. Bibikov, P.P. Kulish,
``Dirac operators on quantum $SU(2)$ group and quantum sphere'',
{\tt q-alg/9608012}


\bibitem{qFSI} H. Grosse, J. Madore, H. Steinacker,
``Field Theory on the q-deformed Fuzzy Sphere I''.
 {\em  J.Geom.Phys.} {\bf 38} (2001) 308-342,
 {\tt hep-th/0005273}

\bibitem{qFSII} H. Grosse, J. Madore, H. Steinacker,
 ``Field Theory on the q-deformed Fuzzy Sphere II: Quantization''.
 {\em  J.Geom.Phys.} {\bf 43} (2002) 205-240,
 {\tt hep-th/0103164}

\bibitem{kirill-resh} A.N. Kirillov, N. Reshetikhin,  "q- Weyl group
  and a Multiplicative Formula for Universal R- Matrices" {\em Comm.
  Math. Phys.} {\bf 134}, 421 (1990)


\bibitem{other}
J. Maldacena, G. Moore, N. Seiberg
``Geometrical interpretation of D-branes in gauged WZW models'',  
{\em  JHEP} {\bf 0107} (2001) 046,  {\tt hep-th/0105038};\\
S. Fredenhagen, V. Schomerus, ``D-Branes in Coset Models'',
 {\em  JHEP} {\bf 0202} (2002) 005, {\tt hep-th/0111189};\\
K. Matsubara, V. Schomerus, M. Smedback
 ``Open Strings in Simple Current Orbifolds'',  
{\em  Nucl.Phys.} {\bf B626} (2002) 53-72, {\tt  hep-th/0108126};\\
S. Fredenhagen, V. Schomerus ``D-Branes in Coset Models'', 
{\em JHEP} {\bf 0202} (2002) 005,   {\tt hep-th/0111189};\\
A. Yu. Alekseev, S. Fredenhagen, T. Quella, V. Schomerus
``Non-commutative gauge theory of twisted D-branes'',  
{\em  Nucl.Phys.} {\bf B646} (2002) 127-157,  {\tt hep-th/0205123}.



\end{thebibliography}
\end{document}